\documentclass[pra,twocolumn,amsmath,nofootinbib,showpacs]{revtex4}
\usepackage{amsmath}
\usepackage{graphicx}
\usepackage{amssymb}
\usepackage{color}

\begin{document}
\input{epsf.sty}

\title{Contextuality for preparations, transformations and unsharp measurements}
\author{R. W. Spekkens}
\email{rspekkens@perimeterinstitute.ca}
\affiliation{Perimeter
Institute for Theoretical Physics, 35 King St. North, Waterloo,
  Ontario N2J 2W9, Canada}
\date{Feb. 25, 2005}

\begin{abstract}
The Bell-Kochen-Specker theorem establishes the impossibility of a
noncontextual hidden variable model of quantum theory, or
equivalently, that quantum theory is contextual. In this paper, an
operational definition of contextuality is introduced which
generalizes the standard notion in three ways: (1) it applies to
arbitrary operational theories rather than just quantum theory,
(2) it applies to arbitrary experimental procedures rather than
just sharp measurements, and (3) it applies to a broad class of
ontological models of quantum theory rather than just
deterministic hidden variable models. We derive three no-go
theorems for ontological models, each based on an assumption of
noncontextuality for a different sort of experimental procedure;
one for preparation procedures, another for unsharp measurement
procedures (that is, measurement procedures associated with
positive-operator valued measures), and a third for transformation
procedures. All three proofs apply to two-dimensional Hilbert
spaces, and are therefore stronger than traditional proofs of
contextuality.
\end{abstract}

\maketitle

\section{Introduction}

Traditionally, a \emph{noncontextual} hidden variable model of
quantum theory is one wherein the measurement outcome that occurs
for a particular set of values of the hidden variables depends
only on the Hermitian operator associated with the measurement and
not on which Hermitian operators are measured simultaneously with
it. For instance, suppose $A$,$B$ and $C$ are Hermitian operators
such that $A$ and $B$ commute, $A$ and $C$ commute, but $B$ and
$C$ do not commute. Then the assumption of noncontextuality is
that the value predicted to occur in a measurement of $A$ does not
depend on whether $B$ or $C$ was measured simultaneously.   The
Bell-Kochen-Specker theorem shows that a hidden variable model of
quantum theory that is noncontextual in this sense is impossible
for Hilbert spaces of dimension three or greater
\cite{Bell,KochenSpecker}.

The traditional definition of noncontextuality is lacking in
several respects: (1) it does not apply to an arbitrary physical
theory, but is rather specific to quantum theory; (2) it does not
apply to unsharp measurements, that is, those associated with
positive-operator valued measures (POVMs), nor does it apply to
preparation or transformation procedures; and (3) it does not
apply to ontological models wherein the outcomes of measurements
are determined only probabilistically from the complete physical
state of the system under investigation, for instance,
indeterministic hidden variable models or ontological models of
quantum theory lacking hidden variables. In this paper, we propose
a new definition:

\begin{quote}
A noncontextual ontological model of an operational theory is one
wherein if two experimental procedures are operationally
equivalent, then they have equivalent representations in the
ontological model.
\end{quote}

This definition will be explained in section
\ref{NCforanyoperationaltheory} of this article, where we provide
a precise account of what it is for two experimental procedures to
be operationally equivalent, and describe what is meant by an
ontological model of an operational theory, specifying in
particular how different experimental procedures (preparations,
measurements and transformations) are represented in such a model.
We also explain why it is appropriate to call this sort of
ontological model \emph{noncontextual} by providing an operational
definition of an experimental context.

In section \ref{NCforquantumtheory}, we specialize our definition
to the case of quantum theory. We provide examples of the sorts of
contexts that can arise for preparations, transformations and
measurements, and describe what an assumption of noncontextuality
for each type of procedure implies for an ontological model of
quantum theory. In the case of measurements, we also generalize
the object that is examined for context-dependence from outcomes
to probabilities of outcomes, and discuss the motivation for doing
so. Further, we show how the traditional notion of
noncontextuality is subsumed as a special case of our generalized
notion when the outcomes of sharp measurements are assumed to be
uniquely determined by the complete physical state of the system
under investigation.

In sections \ref{prepcontextuality}, \ref{BKStheoremin2d}, and
\ref{transfcontextuality}, we provide no-go theorems for
ontological models based on the assumption of noncontextuality for
preparations, unsharp measurements, and transformations,
respectively. All three proofs apply to two-dimensional (2d)
Hilbert spaces, and are therefore stronger than traditional no-go
theorems for noncontextuality, which require Hilbert spaces of
dimension three or greater\footnote{Recent work by Cabello
\cite{Cabello} generalizes the notion of contextuality to unsharp
measurements in a manner that is different from the proposal of
this paper.  From our perspective, this work makes use of an
assumption of deterministic outcomes for unsharp measurements that
cannot be justified by an assumption of noncontextuality.  This
issue is discussed in section \ref{OD}.}. In section
\ref{BKStheoremin2d}, we also provide a no-go theorem for
noncontextuality of unsharp measurements based on a recent
generalization of Gleason's theorem to 2d Hilbert spaces
\cite{Busch,Renes}. Section \ref{isNCnatural} provides a general
discussion of the motivation and plausibility of noncontextuality
for different sorts of procedures, and section \ref{OD}
investigates the connection between these different sorts of
noncontextuality and the assumption that measurement outcomes are
uniquely determined by the complete physical state of the system
under investigation. Some conclusions and questions for future
research are presented in section \ref{conclusions}.

\section{Definitions of noncontextuality for any operational theory}
\label{NCforanyoperationaltheory}

In an operational interpretation of a physical theory, the
primitive elements are preparation procedures, transformation
procedures, and measurement procedures. These are understood as
lists of instructions to be implemented in the laboratory. The
role of an operational theory is merely to specify the
probabilities $p(k|\mathrm{P,T,M})$ of different outcomes $k$ that
may result from a measurement procedure \textrm{M} given a
particular preparation procedure \textrm{P}, and a particular
transformation procedure \textrm{T}. When there is no
transformation procedure, or when it is considered to be part of
the preparation or the measurement, we have simply
$p(k|\mathrm{P,M})$.

Given the rule for determining probabilities of outcomes, one can
define a notion of equivalence among experimental procedures.
Specifically, two preparation procedures are deemed equivalent if
they yield the same long-run statistics for every possible
measurement procedure, that is, $\mathrm{P}$ is equivalent to
$\mathrm{P}^{\prime }$ if
\begin{equation}
p(k|\mathrm{P,M})=p(k|\mathrm{P}^{\prime }\mathrm{,M})\text{ for all }
\mathrm{M.}
\end{equation}
Two measurement procedures are deemed equivalent if they yield the
same long-run statistics for every possible preparation procedure,
that is, $\mathrm{M}$ is equivalent to $\mathrm{M}^{\prime }$ if
their outcomes can be associated one-to-one such that
\begin{equation}
p(k|\mathrm{P,M})=p(k|\mathrm{P,M}^{\prime })\text{ for all }\mathrm{P.}
\end{equation}
Finally, two transformation procedures are deemed equivalent if
they yield the same long-run statistics for every possible
preparation procedure that may precede and every possible
measurement procedure that may follow, that is, $\mathrm{T}$ is
equivalent to $\mathrm{T}^{\prime }$ if
\begin{equation}
p(k|\mathrm{P,T,M})=p(k|\mathrm{P},\mathrm{T}^{\prime }\mathrm{,M})\text{ for all }
\mathrm{P,M.}
\end{equation}

It follows that one can distinguish two types of features of an
experimental procedure: the first type of feature is one that is
specified by specifying the equivalence class that the procedure
falls in, while the second type is one that is not. The set of
features of the second type -- those that are not specified by
specifying the equivalence class -- we call the \emph{context} of
the experimental procedure. Note that by our definition of an
experimental context having knowledge of the context does not
enable one to predict the outcome of an experiment any better than
if one only knew the equivalence class of the experimental
procedure.

An example from quantum theory should clarify the notion of a
context. Consider the following different measurement procedures
for photon polarization. The first, which we denote by
$\mathrm{M}_1$, constitutes a piece of polaroid oriented to pass
light that is vertically polarized along the $\hat{z}$ axis,
followed by a photodetector. The second, which we denote by
$\mathrm{M}_2$, constitutes a birefringent crystal oriented to
separate light that is vertically polarized along the $\hat{z}$
axis from light that is horizontally polarized along this axis,
followed by a photodetector in the vertically polarized output.
The third and fourth procedures, denoted $\mathrm{M}_3$ and
$\mathrm{M}_4$ are identical to $\mathrm{M}_1$ and $\mathrm{M}_2$
respectively, except that they are defined relative to an axis
$\hat{n}$ that is skew to the $\hat{z}$ axis. It turns out that
the statistics of outcomes for $\mathrm{M}_1$ are the same as
those for $\mathrm{M}_2$, for all preparation procedures, and
those for $\mathrm{M}_3$ are the same as those for $\mathrm{M}_4$.
However, the statistics of outcomes for the first pair are
different from those of the second. Thus, $\mathrm{M}_1$ and
$\mathrm{M}_2$ fall in one equivalence class of measurements, and
$\mathrm{M}_3$ and $\mathrm{M}_4$ fall in another. The orientation
of the polaroid or calcite crystal is an example of the first sort
of feature of an experimental operation, one whose variation
involves a variation in the operational equivalence class of the
procedure.  On the other hand, whether one uses a piece of
polaroid or a birefringent crystal to measure photon polarization
is a feature of the measurement procedure of the second type; a
variation of this feature does not change the equivalence class of
the procedure. It is therefore part of the context of the
measurement procedure.

To properly define a noncontextual ontological model of an
operational theory, it is not enough to have a definition of
context; we also need to specify precisely what we mean by an
ontological model. We turn to this now.

An ontological model is an attempt to offer an explanation of the
success of an operational theory by assuming that there exist
physical systems that are the subject of the experiment.  These
systems are presumed to have attributes regardless of whether they
are being subjected to experimental test, and regardless of what
anyone knows about them. These attributes describe the real state
of affairs of the system. Thus, a specification of which instance
of each attribute applies at a given time we call the \emph{ontic
state} of the system. If the ontic state is not completely
specified after specifying the preparation procedure, then the
additional variables required to specify it are called
\emph{hidden variables}. Although most ontological models do
involve hidden variables, this is not always the case. For
instance, the ontic states may be associated one-to-one with the
equivalence classes of preparation procedures (as is the case for
pure preparation procedures in the Beltrametti-Bugajski model of
quantum theory \cite{BeltramettiBugajski}). We shall denote the
complete set of variables in an ontological model by $\lambda$,
and the space of values of $\lambda$ by $\Omega$.

Within an ontological model of an operational theory, preparation
procedures are preparations \emph{of} the ontic state of the
system. However, the procedure need not fix this state uniquely;
rather, it might only fix the probabilities that the system be in
different ontic states. Thus, someone who knows that a system was
prepared using the preparation procedure $\mathrm{P}$ describes
the system by a probability density $\mu _{\mathrm{P}}(\lambda )$
over the model variables.

Similarly, measurement procedures are measurements \emph{of } the
ontic state of the system.  Again, these procedures need not
enable one to infer the identity of the ontic state uniquely, nor
need they even enable one to infer a set of ontic states within
which the actual ontic state lies. Rather, they might only enable
one to infer probabilities for the system to have been in
different ontic states. In this, the most general case, the
outcome of the measurement is not uniquely determined by $\lambda
.$ Only the probabilities of the different outcomes are so
determined. Thus, for every value of $\lambda ,$ one associates a
probability $\xi _{\mathrm{M},k}(\mathbf{\lambda })$ which is the
probability of obtaining outcome $k$ in a measurement $\mathrm{M}$
given that the system is in the ontic state $\lambda .$  We call
$\xi _{\mathrm{M},k}(\mathbf{\lambda })$ an ``indicator
function"\footnote{Some might argue that within the framework of
an ontological model, the term \emph{measurement} ought to be
reserved for a procedure which reveals some attribute of the
system under investigation. However, we feel that it is suitable
for any procedure that leads to an update in one's information
about which instance of some attribute applied, or equivalently,
in one's information about what the ontic state of the system was
prior to the procedure being implemented. Note that even this weak
notion of what constitutes a measurement fails for some
experimental procedures that lead to distinct outcomes, namely,
those wherein the probabilities of the different outcomes are
independent of the ontic state. For simplicity however, we shall
not introduce any novel terminology for this exceptional case.}.

Finally, transformation procedures are transformations \emph{of}
the ontic state of the system. These may be stochastic
transitions. Thus, a transformation procedure $\mathrm{T}$ is
represented by a transition matrix,
$\Gamma_{\mathrm{T}}(\lambda^{\prime},\lambda)$, which represents
the probability density for a transition from the ontic state
$\lambda$ to the ontic state $\lambda^{\prime}$.

Thus, within an ontological model, if the preparation procedure is
$\mathrm{P}$, and the measurement procedure is $\mathrm{M,}$ then
the probability assigned to outcome $k$ is the probability
assigned to outcome $k$ given $\lambda ,$ averaged over all
$\lambda ,$ weighted by the probability of $\lambda ,$
that is, $\int d\mathbf{\lambda }\mu _{\mathrm{P}}(\mathbf{\lambda )}\xi _{%
\mathrm{M},k}(\mathbf{\lambda }).$  If there is a transformation
procedure $\mathrm{T}$ intervening between the preparation and
measurement, and this is associated with the transition matrix
$\Gamma_{\mathrm{T}}(\lambda^{\prime},\lambda)$, then the
probability of outcome $k$ is $\int d\mathbf{\lambda}^{\prime}
d\mathbf{\lambda }\;\xi _{\mathrm{M},k}(\mathbf{\lambda^{\prime}
})\;\Gamma_{\mathrm{T}}(\lambda^{\prime},\lambda)\;\mu
_{\mathrm{P}}(\mathbf{\lambda}) .$

Summarizing, an ontological model assumes: (1) every preparation
procedure \textrm{P} is associated with a normalized probability
density over the ontic state space, $\mu _{\mathrm{P}}: \Omega
\rightarrow [0,1]$ such that $\int \mu _{\mathrm{P}}(\lambda
)d\lambda =1;$ (2) every measurement procedure \textrm{M }with
outcomes labelled by $k$ is associated with a set of indicator
functions $\{\xi _{\mathrm{M},k}(\mathbf{\lambda })\}_{k}$ over
the ontic states, that is, a set of functions $ \xi
_{\mathrm{M},k}:\Omega \rightarrow [0,1]$ satisfying $\sum_{k}\xi
_{\mathrm{M},k}(\mathbf{\lambda })=1$ for all $\lambda$; (3) every
transformation procedure \textrm{T} is associated with a
transition matrix $\Gamma_{\mathrm{T}}:\Omega \times \Omega
\rightarrow [0,1]$ such that $\int
\Gamma_{\mathrm{T}}(\lambda^{\prime},\lambda) d\lambda^{\prime}
=1$ for all $\lambda$, and (4) the predictions of the operational
theory are reproduced exactly by the model, that is,
\begin{equation}
p(k|\mathrm{P,T,M})= \int d\mathbf{\lambda}^{\prime}
d\mathbf{\lambda }\;\xi _{\mathrm{M},k}(\mathbf{\lambda^{\prime}
})\;\Gamma_{\mathrm{T}}(\lambda^{\prime},\lambda)\;\mu
_{\mathrm{P}}(\mathbf{\lambda})
\end{equation}
for all \textrm{P }, \textrm{T}, and \textrm{M. }

In general, the representation of an experimental procedure in an
ontological model might depend on both its equivalence class and
its context.\footnote{If one allows such generality, then --
perhaps contrary to a common impression -- it is possible to
provide an ontological model of quantum theory.  The
deBroglie-Bohm theory \cite{Bohmtheory} is an example.} It is
natural, however, to consider the possibility of a model wherein
the representation of every experimental procedure depends
\emph{only} on its equivalence class and not on its context. After
all, a natural way to explain the fact that a pair of preparation
(measurement,transformation) procedures are operationally
equivalent is to assume that they prepare (measure,transform) the
ontic state of the system in precisely the same way. We shall call
such an ontological model \emph{noncontextual. } Any operational
theory that admits such a model shall also be called
\emph{noncontextual}\footnote{This terminology allows one to use
the phrase "quantum theory is contextual" as a shorthand for
"quantum theory does not admit a noncontextual ontological model",
much as it is common to use the phrase "quantum theory is
nonlocal" in place of "quantum theory does not admit a local
ontological model".}.  In general, if any set of procedures is
represented in a context-independent way within an ontological
model, we shall say that the model is \emph{noncontextual} for
those procedures.

It is useful to explicitly characterize the assumption of
noncontextuality for preparations, transformations, and
measurements. We will call an ontological model \emph{preparation
noncontextual }if the representation of every preparation
procedure is independent of context, that is, if
\begin{equation}
\mu _{\mathrm{P}}(\mathbf{\lambda })=\mu _{e(\mathrm{P})}(\mathbf{\lambda })
\end{equation}
where $e(\mathrm{P})$ is \textrm{P}'s equivalence class.
Similarly, we will call a model \emph{measurement noncontextual
}if the representation of every measurement procedure is
independent of context, that is, if
\begin{equation}
\xi _{\mathrm{M},k}(\mathbf{\lambda })=\xi _{e(\mathrm{M}),k}(\mathbf{%
\lambda })
\end{equation}
where $e(\mathrm{M})$ is \textrm{M}'s equivalence class.  Finally,
a model is called \emph{transformation noncontextual} if the
representation of every transformation procedure is independent of
context, that is, if
\begin{equation}
\Gamma_{\mathrm{T}}(\lambda^{\prime},\lambda)=\Gamma_{e(\mathrm{T})}(\lambda^{\prime},\lambda)
\end{equation}
where $e(\mathrm{T})$ is \textrm{T}'s equivalence class. A
\emph{universally noncontextual} ontological model is one that is
noncontextual for all experimental procedures: preparations,
transformations, and measurements.

\section{Definitions of noncontextuality in quantum theory}
\label{NCforquantumtheory}

We begin with a quick review of the operational approach to
quantum theory, described, for instance, in Refs.\
\cite{Peres,Hardy,Kraus,Buschetal}.

An equivalence class of preparation procedures is associated with
a density operator $\rho $. This is a positive trace-1 operator
over the Hilbert space $\mathcal{H}$ of the system: $\rho > 0$,
$\rm{Tr}(\rho)=1$.  Rank-1 density operators are simply projectors
onto rays of Hilbert space, and are called \emph{pure}.

An equivalence class of measurement procedures is associated with
a positive operator valued measure (POVM) $\{E_{k}\}$.  A POVM is
an ordered set $\{E_{k}\}$ of positive operators that sum to
identity, $\sum_k E_k=I$. The $k$th element, $E_k$, is associated
with the $k$th outcome. Specifically, given a preparation
associated with a density operator $\rho$, the probability of the
outcome $k$ is simply $\rm{Tr}(\rho E_k)$. It is useful to single
out the POVMs whose elements are idempotent, that is, those for
which $E_{k}^2=E_{k}$ for all $k$. Since idempotent positive
operators are projectors, these POVMs are called
\emph{projective-valued} measures (PVMs).  The associated
measurements are said to be \emph{sharp}.  These are the sorts of
measurements that are considered in standard textbook treatments
of quantum mechanics. A Hermitian operator defines a PVM through
the projectors in its spectral resolution.

Finally, an equivalence class of transformation procedures is
associated with a completely positive (CP) map $\mathcal{T}$. A CP
map $\mathcal{T}$ is a positive linear map on the space of
operators over $\mathcal{H}$ such that
$\mathcal{T}\otimes\mathcal{I}$ is a positive linear map on the
space of operators over $\mathcal{H} \otimes
\mathcal{H}^{\prime}$, where $\mathcal{H}^{\prime}$ is of
arbitrary dimension, and $\mathcal{I}$ is the identity map on
$\mathcal{H}^{\prime}$. Unitary maps, familiar from textbook
treatments of quantum theory, are reversible CP maps.

A preparation procedure associated with a non-rank-1 density
operator $\rho $ can be implemented in as many ways as there are
convex decompositions of $\rho$. Suppose $\{(p_{k},\rho _{k})\}$
is a convex decomposition of $\rho ,$ that is, $\rho=\sum_k p_k
\rho_k$.  If one generates a random number according to the
distribution $p_{k},$ and upon obtaining the number $k,$ one
implements the preparation associated with $\rho _{k}$, this
procedure is a member of the equivalence class of procedures
associated with $\rho .$  Another way of implementing a
preparation procedure that is associated with $\rho $ is to
implement a preparation of a purification $\left| \psi
\right\rangle $ of $\rho $ on $\cal{H}\otimes \cal{H}^{\prime }$
(a purification of $\rho $ is any state $\left| \psi \right\rangle
$ such that $\rm{Tr}_{\cal{H}^{\prime }}\left| \psi \right\rangle
\left\langle \psi \right| =\rho )$. The equivalence class
therefore also contains members associated with different
purifications of $\rho .$

The assumption of preparation noncontextuality in quantum theory
is that the probability distribution over ontic states that is
associated with a preparation procedure \textrm{P} depends only on
the density operator $\rho$ associated with \textrm{P},
\begin{equation}
\mu _{\mathrm{P}}(\mathbf{\lambda })=\mu _{\rho }(\mathbf{\lambda }).
\end{equation}
In particular, the distribution does not depend on the particular
convex decomposition of $\rho$ or on the particular purification
of $\rho$ that is used in the preparation procedure.

The multiplicity of contexts for transformation procedures
parallels the multiplicity of contexts for preparation procedures.
Transformation procedures that are associated with non-unitary CP
maps can be obtained as a convex sum of unitary maps in many
different ways, each one of which corresponds to a distinct
transformation procedure, and can also be obtained by implementing
a unitary map on a larger system that incorporates the system of
interest \cite{Schumacher}.

The assumption of transformation noncontextuality in quantum
theory is that the transition matrix that is associated with a
given transformation procedure \textrm{T} depends only on the CP
map $\mathcal{T}$ associated with \textrm{T},
\begin{equation}
\Gamma_{\mathrm{T}}(\lambda^{\prime},\lambda)=\Gamma_{\mathcal{T}}(\lambda^{\prime},\lambda).
\end{equation}
It does not, for instance, depend on the particular convex sum of
unitaries or the particular unitary on a larger system by which
the transformation was implemented.

In the case of quantum measurements, there are also many sorts of
contexts. For instance, every fine-graining of a non-maximally
informative measurement (i.e. a measurement associated with a POVM
at least one element of which is not rank-1) provides a different
context. Suppose the POVM $\{F_j\}$ is a fine-graining of
$\{E_k\}$, which is to say that there is a partitioning of the
outcomes $j$ into sets $S_k$ such that $E_k=\sum_{j\in S_k}F_j$.
By implementing a measurement associated with the POVM $\{F_j\}$,
then discarding all information about $j$ except the set $S_k$ to
which it belongs, one implements a measurement in the equivalence
class associated with the POVM $\{E_k\}$.

Despite the fact that the independence of representation on
fine-graining is traditionally the full extent of the assumption
of noncontextuality for measurements (as we will show below), it
is not difficult to see that there are many other sorts of
contexts. For instance, there is a context for every convex
decomposition of a non-maximal measurement. A convex decomposition
of a POVM $\{E_k\}$ is defined as a probability distribution
$\{p_{\alpha}\}$ and a set of POVMs, $\{\cal{F}_{\alpha}\}$ where
${\cal F}_{\alpha}=\{F_k^{\alpha}\}$, such that $E_k=\sum_{\alpha}
p_{\alpha}F_k^{\alpha}$. By sampling $\alpha$ from the
distribution $\{p_{\alpha}\}$, then implementing the measurement
associated with the POVM $\{F_k^{\alpha}\}$, and registering only
the outcome of this measurement, one implements a measurement in
the equivalence class associated with the POVM $\{E_k\}$. There is
also a context for every way of obtaining a POVM by coupling to an
ancilla and measuring a PVM on the composite of system+ancilla
\cite{PeresNeumark}.

The assumption of measurement noncontextuality in quantum theory
is that the set of indicator functions representing a measurement
\textrm{M }depends only on the POVM $\{E_{k}\}$ associated with
\textrm{M,}
\begin{equation}
\xi _{\mathrm{M},k}(\mathbf{\lambda })=\xi _{\{E_{k}\},k}(\mathbf{\lambda }%
).
\end{equation}

In addition to admitting new sorts of contexts for measurements,
our generalized notion of measurement contextuality involves a
slight revision of \emph{what it is} that depends on the
measurement context.

In the past, measurement contextuality has only been considered
within the framework of deterministic hidden variable theories,
and the question of interest has been whether or not the
measurement outcome for a given ontic state of the system depends
on the context of the measurement. However, for \emph{objectively
indeterministic} ontological models, it is clear that the natural
question to ask is whether the \emph{probabilities} of different
outcomes for a given ontic state of the system depend on the
context. This is analogous to Bell's \cite{Belllocalbeables}
generalization of the notion of locality from measurement outcomes
being causally independent of parameter settings at space-like
separation to the \emph{probabilities} of measurement outcomes
being causally independent of parameter settings at space-like
separation\footnote{\label{footnotelocality}More specifically,
Bell \cite{Belllocalbeables} defined a theory to be \emph{locally
deterministic} if the variables in space-time region I are
determined by the variables in a space-time region that fully
closes the backward light-cone of I, and \emph{locally causal} if
the probability distribution over values for a variable in
space-time region I are determined by a specification of the
values of all the variables in the backward light-cone of I
(``determined" in the sense that further conditioning on variables
in the region outside the backward light-cone would not change the
probability distribution).}.  This distinction was introduced by
Bell in order to cleanly separate the notion of locality from the
notion of determinism.  Similarly, our generalized definition
allows one to cleanly separate the notion of measurement
noncontextuality from the notion of determinism.

Once the question is posed, it is somewhat obvious that if there
is to be any notion of measurement contextuality within
objectively indeterministic ontological models, the appropriate
quantities to examine for context-dependence are the
\emph{probabilities} of different outcomes for a given ontic state
of the system.  A less obvious feature of our generalized notion
of measurement contextuality is that the probabilities of outcomes
are the appropriate quantities to examine for context-dependence
even in \emph{objectively deterministic} ontological models. The
key is that the latter sort of model may still exhibit an
\emph{epistemic} indeterminism, wherein knowledge of the
equivalence class of the measurement together with the ontic state
of the system under investigation does not uniquely fix the
outcome.  To explain this properly, we need to consider some of
the details of the mathematical representation of these
measurements.

The distinction between the ontic state of the system determining
the outcome and determining only the probabilities of different
outcomes is captured mathematically within an ontological model by
the sorts of indicator functions one uses to represent
measurements. The former case is represented by an indicator
function that is \emph{idempotent}, that is, one for which
$\chi(\lambda)^2=\chi(\lambda)$ (we shall denote idempotent
indicator functions by $\chi(\lambda)$ rather than
$\xi(\lambda)$). Such functions are necessarily equal to one in
some region of the ontic state space and zero elsewhere. By virtue
of the fact that a set of indicator functions must satisfy $\sum_k
\chi _{k}(\lambda)=1$, if all the indicator functions are
idempotent, then the latter must be nonoverlapping, that is, $\chi
_{k}(\mathbf{\lambda })\chi _{k^{\prime }}(\mathbf{\lambda })=0$
for $k\ne k^{\prime }.$ Thus, for every value of $\lambda ,$ only
a single indicator function in the set $\{\chi _{k}(\lambda )\}$
receives the value $1$ while the others receive the value $0.$
Since the value of the $k$th indicator function at a given
$\lambda$ specifies the probability of the $k$th outcome given the
ontic state $\lambda$ , the outcome of the measurement is
determined for all ontic states if and only if the latter is
represented by a set of idempotent indicator functions.

We shall call the assumption that a particular measurement is
represented by a set of idempotent indicator functions the
assumption of \emph{outcome determinism }for that measurement.

Now note that even within an objectively deterministic ontological
model, measurements may fail to exhibit outcome determinism:
specifying the ontic state of the system under investigation
together with the equivalence class of the measurement procedure
may be insufficient to uniquely fix the outcome.  The outcome
might only be fixed uniquely by supplementary features of the
measurement procedure (which constitute part of the context of the
measurement by our definition), such as microscopic degrees of
freedom of the apparatus. Because the indicator function for a
measurement specifies the dependence of the outcome on the ontic
state of the system under investigation, and not the dependence of
the outcome on the ontic state of any systems that make up the
measurement apparatus or the environment, such a measurement must
be represented by a non-idempotent set of indicator functions.
Nonetheless, it may still be the case that for each equivalence
class of measurements, all the elements of the class are
represented by the \emph{same} non-idempotent set of indicator
functions, and \emph{this} is all that is required for the
measurements to be deemed noncontextual by our definition.

As an example, consider a classical system and a classical
measurement device that generates an outcome by rolling one of
several differently weighted dice, with the choice of the dice
being determined by the ontic state of the system. Two such
devices are only found to be operationally equivalent if all of
the dice of one are weighted in the same way as those of the
other.  Thus, every device in the equivalence class is represented
by the same set of indicator functions, and consequently one has
measurement noncontextuality by our definition. The underlying
ontological model (classical mechanics) is objectively
deterministic, but in order to predict the outcome of a particular
measurement, one must supplement the ontic state of the system by
the precise initial configuration of the dice and their
environment, features that form part of the context of the
measurement. Thus, although the outcome of the measurement clearly
depends on the context, we take this to be a failure of outcome
determinism rather than a failure of measurement
noncontextuality\footnote{By our definition, any classical theory
is necessarily noncontextual for all experimental procedures. This
highlights another virtue of our particular definition:
contextuality, in all of its manifestations, is found to be a
nonclassical phenomenon. For an opposing perspective, see Ref.
\cite{Bacciagaluppi}.}.

Thus, it is really the notion of \emph{outcome determinism},
rather than the notion of determinism, which we seek to cleanly
separate from the notion of measurement noncontextuality through
our generalized definition. This makes our definition of
measurement noncontextuality revisionist insofar as the
traditional definition implicitly incorporated the assumption of
outcome determinism, while ours does not.  This suggested revision
in terminology is motivated by the idea that what is crucial to
the notion of a noncontextual ontological model is that it
reproduces the equivalence class structure of the operational
theory.

It is worth noting that, given the additional assumption of
outcome determinism, one can recover the traditional definition of
measurement noncontextuality as a special case of our definition.
Specifically, if one considers only sharp measurements and one
represents these by sets of idempotent functions (i.e. one assumes
outcome determinism for these measurements), then the assumption
of the independence of the representation of a measurement on the
fine-graining of the PVM with which it is implemented is just the
traditional notion of noncontextuality (described in the
introduction). This can be seen as follows. Specifying whether a
Hermitian operator $A$ is measured together with $B$ or with $C$
is equivalent to specifying a fine-graining of the PVM $\{P_k\}$
that is defined by the spectral resolution of $A$; the
simultaneous eigenspaces of $A$ and $B$ define one such
fine-graining, while the simultaneous eigenspaces of $A$ and $C$
define another. Specifying the eigenvalue assigned to an operator
$A$ for every value of $\lambda$ is equivalent to specifying a set
of idempotent indicator functions; the values of $\lambda$ in the
support of the function associated with outcome $k$ are simply
those that assign the $k$th eigenvalue to $A$.  Clearly then,
assuming that the value assigned to $A$ is independent of whether
$A$ is measured together with $B$ or $C$ is equivalent to assuming
that the set of idempotent indicator functions associated with the
PVM $\{P_k\}$ is independent of the fine-graining by which it was
implemented.

No-go theorems based on the traditional definition of
noncontextuality apply only in Hilbert spaces of dimensionality
three or greater.  Moreover, one cannot extend such proofs to 2d
Hilbert spaces because there are no fine-grainings of non-trivial
PVM measurements in a 2d Hilbert space, and fine-graining is the
only notion of context that is recognized traditionally. However,
by appealing to preparations, transformations, and unsharp
measurements, which admit many contexts even in a 2d Hilbert
space, proofs of contextuality can be achieved here as well. From
this perspective, the restriction of previous proofs of
contextuality to 3d Hilbert spaces was an artifact of a limited
notion of a context.

Among the new proofs of contextuality that we shall present, the
proof of preparation contextuality is the simplest, and so we
begin with this case.

\section{Proof of preparation contextuality in 2D} \label{prepcontextuality}

There are two features of the representation of preparation
procedures in an ontological model that are central to our proof.
The first concerns distinguishability, and the second convex
combination.\medskip

\noindent\textit{Feature 1} If two preparation procedures, P and
P$'$ are distinguishable with certainty in a single-shot
measurement, then their associated probability distributions,
$\mu(\lambda)$ and $\mu'(\lambda)$, are nonoverlapping, that is,
\begin{equation}
\mu(\mathbf{\lambda )} \mu'(\mathbf{\lambda })=0 \text{ for all
}\lambda. \label{disjointness}
\end{equation}

This feature can be understood as follows. Suppose one wishes to
perform a measurement that discriminates, with certainty, between
two probability distributions. In other words, one wishes to
perform a measurement that allows one to retrodict, with
certainty, which distribution applied. This is only possible if
the distributions to be discriminated are nonoverlapping. The
reason is that if the two distributions overlapped in some region
of the space of ontic states, then whenever the actual ontic state
was in that region (and it will sometimes be in that region,
because the region is assigned nonzero probability by both
distributions), no measurement would be able to distinguish with
certainty whether the system had been prepared using one or the
other distribution, since the actual ontic state is consistent
with both. Thus, if, within an operational theory, a pair of
preparation procedures are distinguishable with certainty, then
the only way an ontological model of the theory can account for
this fact is by associating these procedures with nonoverlapping
distribution functions.

The second feature of an ontological model that is critical to our
proof is the manner in which convex combinations of preparation
procedures are represented. Suppose that the preparation
procedures $\mathrm{P}$ and $\mathrm{P^{\prime}}$ are represented
by distributions $\mu(\lambda)$ and $\mu^{\prime}(\lambda)$. Now
suppose that a bit is generated uniformly at random from the
distribution ${p,1-p}$, and the value of the bit is used to
determine whether $\mathrm{P}$ or $\mathrm{P}^{\prime}$ is
implemented, after which the bit is forgotten.  This effective
procedure, which we call $\mathrm{P}^{\prime \prime}$, must be
represented within the ontological model by a distribution
$\mu^{\prime \prime}(\lambda)$ satisfying
\begin{equation} \mu^{\prime
\prime}(\lambda)=p\mu(\lambda) + (1-p)\mu^{\prime}(\lambda).
\end{equation}
The reason is as follows. The probability that the ontic state of
the system is $\lambda$ given procedure $\mathrm{P}^{\prime
\prime}$, is simply the sum of the probability that it is
$\lambda$ given procedure $\mathrm{P}$ and the probability that it
is $\lambda$ given procedure $\mathrm{P}'$, weighted by the
respective probabilities of $\mathrm{P}$ and $\mathrm{P}'$ given
$\mathrm{P}^{\prime \prime}$.

Thus, we have:\medskip

\noindent\textit{Feature 2} A convex combination of preparation
procedures is represented within an ontological model by a convex
sum of the associated probability distributions.\medskip

With these facts in hand, we now proceed with the proof.

Consider a set of six pure preparations, denoted
$\mathrm{P}_{a},\mathrm{P}_{A},\mathrm{P}_{b},\mathrm{P}_{B},\mathrm{P}_{c},$
and $\mathrm{P}_{C},$ corresponding to the normalized Hilbert
space vectors
\begin{eqnarray}
\psi _{a} &=&(1,0)  \nonumber \\
\psi _{A} &=&(0,1)  \nonumber \\
\psi _{b} &=&(1/2,\sqrt{3}/2)  \nonumber \\
\psi _{B} &=&(\sqrt{3}/2,-1/2)  \nonumber \\
\psi _{c} &=&(1/2,-\sqrt{3}/2)  \nonumber \\
\psi _{C} &=&(\sqrt{3}/2,1/2)  \label{states}
\end{eqnarray}
or, equivalently, the rank 1 density operators
\begin{eqnarray}
\sigma _{a} &=&\left(
\begin{array}{cc}
1 & 0 \\
0 & 0
\end{array}
\right) \nonumber \\
\sigma _{A} &=&\left(
\begin{array}{cc}
0 & 0 \\
0 & 1
\end{array}
\right) \nonumber \\
\sigma _{b} &=&\left(
\begin{array}{cc}
\frac{1}{4} & \frac{1}{4}\sqrt{3} \\
\frac{1}{4}\sqrt{3} & \frac{3}{4}
\end{array}
\right) \nonumber \\
\sigma _{B} &=&\left(
\begin{array}{cc}
\frac{3}{4} & -\frac{1}{4}\sqrt{3} \\
-\frac{1}{4}\sqrt{3} & \frac{1}{4}
\end{array}
\right) \nonumber \\
\sigma _{c} &=&\left(
\begin{array}{cc}
\frac{1}{4} & -\frac{1}{4}\sqrt{3} \\
-\frac{1}{4}\sqrt{3} & \frac{3}{4}
\end{array}
\right) \nonumber \\
\sigma _{C} &=&\left(
\begin{array}{cc}
\frac{3}{4} & \frac{1}{4}\sqrt{3} \\
\frac{1}{4}\sqrt{3} & \frac{1}{4}
\end{array}
\right) \label{states2}
\end{eqnarray}
\newline
One can easily verify the following orthogonality conditions
\begin{eqnarray}
\sigma _{a}\sigma _{A} &=&0,  \label{orthogonality1} \\
\sigma _{b}\sigma _{B} &=&0,  \label{orthogonality2} \\
\sigma _{c}\sigma _{C} &=&0.  \label{orthogonality3}
\end{eqnarray}

Now consider the preparation procedure wherein one of
$\mathrm{P}_a$ or $\mathrm{P}_A$ is implemented, with the choice
being made uniformly at random (for instance, by flipping a fair
coin), and with no record being made of the choice. Denote this
procedure by $\mathrm{P}_{aA}$.  Define procedures
$\mathrm{P}_{bB}$ and $\mathrm{P}_{cC}$ similarly. Consider also
the preparation procedure wherein one of $\mathrm{P}_a$,
$\mathrm{P}_b$, or $\mathrm{P}_C$ is implemented, again, with
equal probabilities for each, and without recording the choice. We
denote this by $\mathrm{P}_{abc}$.  The procedure
$\mathrm{P}_{ABC}$ is defined similarly.

These procedures are represented in the quantum formalism by the
appropriate convex sums of the density operators in Eq.\
(\ref{states2}).  It turns out that all of these convex sums yield
the same rank-2 density operator, namely,
\begin{equation}
I/2=\left(
\begin{array}{cc}
\frac{1}{2} & 0 \\
0 & \frac{1}{2}
\end{array}
\right) ,
\end{equation}
commonly referred to as the `completely mixed state'.
Specifically,
\begin{eqnarray}
I/2 &=&\frac{1}{2}\sigma _{a}+\frac{1}{2}\sigma _{A}  \label{convex1} \\
&=&\frac{1}{2}\sigma _{b}+\frac{1}{2}\sigma _{B}  \label{convex2} \\
&=&\frac{1}{2}\sigma _{c}+\frac{1}{2}\sigma _{C}  \label{convex3} \\
&=&\frac{1}{3}\sigma _{a}+\frac{1}{3}\sigma _{b}+\frac{1}{3}\sigma
_{c}
\label{convex4} \\
&=&\frac{1}{3}\sigma _{A}+\frac{1}{3}\sigma _{B}+\frac{1}{3}\sigma
_{C}. \label{convex5}
\end{eqnarray}

\begin{figure}[h]
\includegraphics[width=40mm]{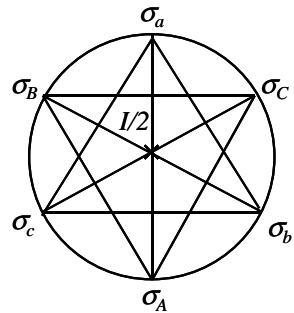}
\caption{The Bloch ball representation of the six pure states and
the five convex decompositions of the completely mixed state used
in the proof of preparation contextuality.  Each convex
decomposition is represented by a convex polytope whose vertices
represent the elements of the decomposition
\cite{SpekkensRudolphQIC}. The 2-element decompositions in our
example are represented by line segments, and the 3-element
decompositions by equilateral triangles. } \label{Bloch1}
\end{figure}

In figure \ref{Bloch1}, we present the Bloch ball representation
of the seven density operators defined above.  This provides a
graphical synopsis of the relevant orthogonality relations and
convex structure.

Within an ontological model of operational quantum theory, each
preparation procedure $\mathrm{P}_x$ is associated with a
probability distribution $\mu_x(\lambda)$. Now note that if two
density operators, $\sigma$ and $\sigma'$, are orthogonal in the
vector space of operators, that is, $\sigma \sigma' = 0$, then the
associated preparation procedures can be distinguished with
certainty in a single-shot measurement (for instance, for
preparations associated with orthogonal Hilbert space vectors, one
simply implements the measurement associated with an orthogonal
basis that includes these vectors). By feature 1 of ontological
models (described above), distinguishable procedures are
represented by nonoverlapping distributions.  Thus, from Eqs.\
(\ref{orthogonality1})-(\ref{orthogonality3}) we can infer that
\begin{eqnarray}
\mu _{a}(\mathbf{\lambda})\mu _{A}(\mathbf{\lambda}) &=&0,  \label{c4} \\
\mu _{b}(\mathbf{\lambda})\mu _{B}(\mathbf{\lambda}) &=&0,  \label{c5} \\
\mu _{c}(\mathbf{\lambda})\mu _{C}(\mathbf{\lambda}) &=&0.
\label{c6}
\end{eqnarray}

Furthermore, in any ontological model a convex combination of
preparation procedures is represented by a convex sum of the
associated probability distributions (feature 2 above).  Thus, if
the procedures
$\mathrm{P}_{aA},\mathrm{P}_{bB},\dots,\mathrm{P}_{ABC}$ are
represented by distributions
$\mu_{aA}(\lambda),\mu_{bB}(\lambda),\dots,\mu_{ABC}(\lambda)$,
the manner in which these procedures are obtained by convex
combination of
$\mathrm{P}_{a},\mathrm{P}_{b},\dots,\mathrm{P}_{C}$ implies that
\begin{eqnarray}
\mu_{aA} (\mathbf{\lambda }) &=&\frac{1}{2}\mu
_{a}(\mathbf{\lambda })+\frac{1}{2}
\mu _{A}(\mathbf{\lambda })  \label{e1} \\
\mu_{bB} (\mathbf{\lambda }) &=&\frac{1}{2}\mu
_{b}(\mathbf{\lambda })+\frac{1}{2}\mu
_{B}(\mathbf{\lambda})  \label{e2} \\
\mu_{cC} (\mathbf{\lambda }) &=&\frac{1}{2}\mu
_{c}(\mathbf{\lambda })+\frac{1}{2}\mu _{C}(\mathbf{\lambda})
 \label{e3} \\
\mu_{abc} (\mathbf{\lambda }) &=&\frac{1}{3}\mu
_{a}(\mathbf{\lambda})+\frac{1}{3}\mu
_{b}(\mathbf{\lambda})+\frac{1}{3}\mu _{c}(\mathbf{\lambda})  \label{e4} \\
\mu_{ABC} (\mathbf{\lambda }) &=&\frac{1}{3}\mu
_{A}(\mathbf{\lambda})+\frac{1}{3}\mu
_{B}(\mathbf{\lambda})+\frac{1}{3}\mu _{C}(\mathbf{\lambda}).
\label{e5}
\end{eqnarray}

The assumption of a preparation noncontextual ontological model is
that the distribution associated with a preparation procedure
depends only on the operational equivalence class of that
procedure, and thus only on the density operator associated with
that procedure. Since the procedures
$\mathrm{P}_{aA},\mathrm{P}_{bB},\dots,\mathrm{P}_{ABC}$ are all
represented by $I/2$, they must all be represented by the same
distribution in a preparation noncontextual ontological model.
Thus, we require $\mu_{aA}=\mu_{bB}=\dots=\mu_{ABC}$.  Denoting
this distribution by $\nu(\lambda)$, we have simply
\begin{eqnarray}
\nu (\mathbf{\lambda }) &=&\frac{1}{2}\mu _{a}(\mathbf{\lambda
})+\frac{1}{2}
\mu _{A}(\mathbf{\lambda })  \label{c7} \\
&=&\frac{1}{2}\mu _{b}(\mathbf{\lambda })+\frac{1}{2}\mu
_{B}(\mathbf{\lambda})  \label{c8} \\
&=&\frac{1}{2}\mu _{c}(\mathbf{\lambda })+\frac{1}{2}\mu
_{C}(\mathbf{\lambda})  \label{c9} \\
&=&\frac{1}{3}\mu _{a}(\mathbf{\lambda})+\frac{1}{3}\mu
_{b}(\mathbf{\lambda})+\frac{1}{3}\mu _{c}(\mathbf{\lambda})  \label{c10} \\
&=&\frac{1}{3}\mu _{A}(\mathbf{\lambda})+\frac{1}{3}\mu
_{B}(\mathbf{\lambda})+\frac{1}{3}\mu _{C}(\mathbf{\lambda}).
\label{c11}
\end{eqnarray}

We now show that there is no set of distributions satisfying Eqs.\
(\ref{c4})-(\ref{c6}) and Eqs.\ (\ref{c7})-(\ref{c11}). Consider
the values of the various probability densities at a fixed value
of $\mathbf{\lambda .}$ We denote these simply as $\mu _{a},\mu
_{A},\dots,\mu _{C}.$ We show that the only solution to all the
constraints, for a fixed $\mathbf{\lambda },$ is $\mu _{a},\mu
_{A},\dots ,\mu _{C}=0,$ which we call the all-zero solution.

To satisfy Eqs.\ (\ref{c4})-(\ref{c6}), one of the pair $\mu _{a}$
and $\mu _{A}$ must be zero, as must be one of the pair $\mu _{b}$
and $\mu _{B}$ and one of the pair $\mu _{c}$ and $\mu _{C}.$ In
all, there are eight possible assignments of zeroes that satisfy
Eqs.\ (\ref{c4})-(\ref{c6}). We consider each of these in turn.

\strut If we have $\mu _{a},\mu _{b},\mu _{c}=0$ then by Eq.\
(\ref{c10}) we have $\nu =0$, and by Eqs.\ (\ref{c7})-(\ref{c9}),
we conclude that $\mu _{A},\mu _{B},\mu _{C}=0,$ so that we have
the all-zero solution. If, instead we have $\mu _{a},\mu _{b},\mu
_{C}=0$ then by combining Eq.\ (\ref {c9}) and Eq.\ (\ref{c10}) we
find $\frac{1}{2}\mu _{c}=\frac{1}{3}\mu _{c},$ for which the only
solution is $\mu _{c}=0.$ But this gets us back to the first case,
and the all-zero solution. Every other case yields the all-zero
solution by virtue of the symmetry of the problem under rotations
by multiples of 60 degrees in the Bloch sphere representation.

The above argument did not depend on $\mathbf{\lambda ,}$ and thus for all $%
\mathbf{\lambda }$ the only solution is the all-zero solution.
Consequently, the only set of distributions that satisfy Eqs.\
(\ref{c4})-(\ref{c6}) and Eqs.\ (\ref{c7})-(\ref{c11}) is the set
of uniformly zero distributions, $\mu _{a}(\mathbf{\lambda }),\mu
_{A}( \mathbf{\lambda }),\cdots ,\mu _{C}(\mathbf{\lambda })=0$.
But such distributions are not probability distributions since
they are not normalized to one. This concludes the proof.

I am grateful to Terry Rudolph for having improved upon my
original proof of preparation contextuality by proposing the
highly symmetric example presented here.

\section{Proofs of contextuality for unsharp measurements in 2d}
\label{BKStheoremin2d}

Proofs of measurement contextuality have usually arisen only in
the context of \emph{sharp} measurements, that is, those
associated with PVMs\footnote{Exceptions are Refs.\
\cite{Cabello,TonerBacon,Bacciagaluppi}}, and outcome determinism
has been assumed for such measurements.  We shall make the same
assumption here for sharp measurements, but we shall be
considering unsharp measurements as well, that is, those
associated with POVMs, and for these, outcome determinism will not
be assumed. It is important to note that the ``proofs of
contextuality" presented in the next two subsections are
contingent on the assumption of outcome determinism for sharp
measurements.  The status of this assumption will be revisited in
section \ref{OD}, where we will clarify what, precisely, has been
proven.

\subsection{A proof based on a finite set of measurements}

Consider three binary-outcome measurements,
$\mathrm{M}_{a},\mathrm{M}_{b}$, and $\mathrm{M}_{c},$ associated
respectively with PVMs $ \{P_{a},P_{A}\},$ $\{P_{b},P_{B}\}$ and
$\{P_{c},P_{C}\},$ where $P_{a}$ projects onto the ray spanned by
$\psi _{a},$ $P_{A}$ projects onto the ray spanned by $\psi _{A},$
and so forth, with the vectors $\psi _{x}$ being those that are
defined in Eq.\ (\ref{states})\footnote{Note that $P_{a}=\sigma
_{a},$ $P_{A}=\sigma _{A},$ etcetera, where $ \sigma _{x}$ is
defined in Eq.\ (\ref{states2}). This follows from the fact that
the rank-1 density operator associated with a vector is simply the
projector onto the ray spanned by that vector.  It follows that
Eqs.\ (\ref{mm1})-(\ref{mm3}) and Eqs.\ (\ref{mm4})-(\ref{mm6})
are equivalent to Eqs.\ (\ref{convex1})-(\ref{convex3}) and Eqs.\
(\ref{orthogonality1})-(\ref{orthogonality3}) respectively. We use
a distinct notation for the same mathematical operators to remind
the reader of the fact that in this section they represent
measurement outcomes rather than preparation procedures.}.

By the definition of a PVM, we have
\begin{eqnarray}
&&P_{a}+P_{A}=I, \label{mm1} \\
&&P_{b}+P_{B}=I, \label{mm2} \\
&&P_{c}+P_{C}=I, \label{mm3}
\end{eqnarray}
and
\begin{eqnarray}
&&P_{a}P_{A}=0, \label{mm4} \\
&&P_{b}P_{B}=0, \label{mm5} \\
&&P_{c}P_{C}=0. \label{mm6}
\end{eqnarray}

Given the assumption of outcome determinism for sharp
measurements,
the representations of $\mathrm{M}_{a},\mathrm{M}_{b}$, and $%
\mathrm{M}_{c}$ in an ontological model are the sets of idempotent
indicator functions $\{\chi _{a}(\lambda ),\chi _{A}(\lambda
)\},\{\chi _{b}(\lambda ),\chi _{B}(\lambda )\},$ and $\{\chi
_{c}(\lambda ),\chi _{C}(\lambda )\}$ respectively. By definition,
these must satisfy
\begin{eqnarray}
&&\chi _{a}(\lambda )+\chi _{A}(\lambda )=1,  \label{m1} \\
&&\chi _{b}(\lambda )+\chi _{B}(\lambda )=1,  \label{m2} \\
&&\chi _{c}(\lambda )+\chi _{C}(\lambda )=1,  \label{m3}
\end{eqnarray}
and
\begin{eqnarray}
\chi _{a}(\lambda )\chi _{A}(\lambda ) &=&0,  \label{m4} \\
\chi _{b}(\lambda )\chi _{B}(\lambda ) &=&0,  \label{m5} \\
\chi _{c}(\lambda )\chi _{C}(\lambda ) &=&0.  \label{m6}
\end{eqnarray}

Now consider choosing one of $\mathrm{M}_{a},\mathrm{M}_{b}$, and
$\mathrm{M}_{c}$ at random, with probability 1/3 for each,
implementing the chosen measurement, and only registering whether
the first (small letter) or the second (capital letter) outcome
occurred. Call the effective measurement procedure that results
$\mathrm{M}$. It is associated with the POVM
\begin{equation}
\{\tfrac{1}{3}P_{a}+\tfrac{1}{3}P_{b}+\tfrac{1}{3}P_{c},\tfrac{1}{3}P_{A}+%
\tfrac{1}{3}P_{B}+\tfrac{1}{3}P_{C}\}. \label{binaryPOVM}
\end{equation}

In an ontological model, a convex combination of measurements
procedures is represented by an element-wise convex sum of the
associated sets of indicator functions (for the same reason that
an ontological model has feature 2 of section
\ref{prepcontextuality}). Thus, $\mathrm{M}$ is represented by the
set of indicator functions
\begin{eqnarray}
\{\tfrac{1}{3}\chi _{a}(\lambda )+\tfrac{1}{3}\chi _{b}(\lambda
)+\tfrac{1}{ 3 }\chi _{c}(\lambda ), \nonumber \\
 \tfrac{1}{3}\chi
_{A}(\lambda )+\tfrac{1}{3}\chi _{B}(\lambda )+\tfrac{1}{3}\chi
_{C}(\lambda )\}. \label{indicatorset1}
\end{eqnarray}

Note that the POVM (\ref{binaryPOVM}) is equal to\footnote{This
fact is also captured by Eqs.\ (\ref{convex4}) and
(\ref{convex5}).}
\begin{equation}
\{\tfrac{1}{2}I,\tfrac{1}{2}I\}. \label{trivialPOVM}
\end{equation}

But it is clear from this way of writing the POVM that the
measurement has a random outcome regardless of the preparation
procedure, since Tr$(\rho \tfrac{1}{2}I)=\tfrac{1}{2}$ regardless
of $\rho$. It then follows that the equivalence class of
measurement procedures that contains $\mathrm{M}$ also contains
the "measurement" procedure $\mathrm{ \tilde{M}}$ that completely
ignores the system and just flips a fair coin to determine the
outcome. Now consider how the measurement $\mathrm{\tilde{M}}$ is
represented in the ontological model. Because the outcome doesn't
depend on the system at all, it follows that regardless of the
value of $\lambda $, there is a probability of $1/2$ for each
outcome, so it is represented by the set of indicator functions
\begin{equation}
\{\tfrac{1}{2},\tfrac{1}{2}\}, \label{indicatorset2}
\end{equation}
 where each element should be thought of as a
uniform function over $\lambda $ of height ${\textstyle {\frac{1
}{2}}}$\footnote{\label{footnoteonpermutation} This fact can also
be established by noting that the equivalence class includes the
measurement $\mathrm{M}^{\prime },$ obtained from $ \mathrm{M}$ by
permuting the two outcomes (because such a permutation does not
change the statistics of outcomes). The ontological
representations of $\mathrm{M}$ and $\mathrm{M}^{\prime }$ are
$\{\xi _{1}(\lambda ),\xi _{2}(\lambda )\}$ and $\{\xi
_{2}(\lambda ),\xi _{1}(\lambda )\}.$ Now, the assumption of
measurement noncontextuality implies that since $\mathrm{M}$ and
$\mathrm{ M}^{\prime }$ are in the same equivalence class, they
must be represented by the same set of indicator functions. Thus
we require that $\xi _{1}(\lambda )=\xi _{2}(\lambda ).$ But since
$\xi _{1}(\lambda )+\xi _{2}(\lambda )=1,$ it follows that $\xi
_{1}(\lambda )=\xi _{2}(\lambda )=1/2$ for all $\lambda. $}.

By the assumption of measurement noncontextuality, the measurement
$\mathrm{M}$ must be represented by the same set of indicator
functions as the measurement $ \mathrm{\tilde{M}}.$ It follows
that the set of functions (\ref{indicatorset1}) must be equal to
the set of functions (\ref{indicatorset2}). However, this
constraint is inconsistent with the constraints
(\ref{m1})-(\ref{m6}). To satisfy Eqs.\ (\ref{m1})-(\ref{m6}) it
is necessary that for every value of $\lambda ,$ one of $\chi
_{a}(\lambda )$ and $\chi _{A}(\lambda )$ must be equal to 0 and
the other equal to $1.$ The same is true of $\chi _{b}(\lambda )$
and $\chi _{B}(\lambda )$ and of $\chi _{c}(\lambda )$ and $\chi
_{C}(\lambda ).$ The eight possible assignments of values to these
six quantities leave the set of functions (\ref{indicatorset1})
with the values $\{0,1\},\{1,0\},\{\tfrac{2}{3},\tfrac{1}{3}\}$ or
$\{\tfrac{1}{3},\tfrac{2}{3}\}$ but never
$\{\tfrac{1}{2},\tfrac{1}{2}\}.$ This concludes the proof.

\subsection{A proof based on the 2D version of Gleason's theorem}

The impossibility of noncontextuality for unsharp measurements and
outcome determinism for sharp measurements can also be established
in a 2d Hilbert space by making appeal to a recent Gleason-like
derivation of the quantum probability rule by Busch \cite{Busch}
and by Caves \emph{et al.} \cite{Renes}. This ``generalized
Gleason's theorem'' starts from the assumption that there exists a
probability measure that assigns a unique probability $w(E)$ to
every positive operator $E$ such that $w(I)=1,$ and whenever a set
of positive operators forms a resolution of identity,
$\sum_{k}E_{k}=I,$ the associated probabilities sum to $1,$
$\sum_{k}w(E_{k})=1.$ From these assumptions, it is proven that
the measure must satisfy $w(E)=\mathrm{Tr}(\rho E)$ for some
density operator $\rho$ \cite{Busch,Renes}.

Recall that the values of a set of indicator functions
$\{\xi_k(\lambda)\}$ at a particular value of $\lambda$ form a
probability distribution over $k$. In a measurement noncontextual
theory, every positive operator $E$ is represented by a unique
indicator function $\xi_E(\lambda)$, with the identity operator
being represented by the unit function. Moreover, whenever a set
of positive operators forms a resolution of identity,
$\sum_{k}E_{k}=I,$ the associated indicator functions sum to the
unit function, $\sum_{k}\xi_{E_k}(\lambda)=1.$  Thus, the set of
all indicator functions for a given value of $\lambda$ in an
ontological model satisfy the assumptions of the set of
probability measures in the generalized Gleason's theorem.  It
follows therefore that for every value of $\lambda$ in the
ontological model, there is a density operator $\rho_{\lambda}$
such that $\xi_E(\lambda)= \mathrm{Tr}(\rho_{\lambda} E)$.

If in addition to measurement noncontextuality, one assumes
outcome determinism for sharp measurements, then every projector
is represented by a unique idempotent indicator function $\chi_P
(\lambda)$, and by the generalized Gleason's theorem,
\begin{equation}
\label{Gleasonconstraint}
\chi_P(\lambda)=\mathrm{Tr}(\rho_{\lambda} P).
\end{equation}

Suppose that $P=\left| \psi \right\rangle \left\langle
\psi\right|$, and consider a $\lambda$ such that $\chi_P
(\lambda)=1$. In this case, Eq.\ (\ref{Gleasonconstraint}) implies
that $\rho_{\lambda}=\left| \psi \right\rangle \left\langle
\psi\right|$.  But then for some other projector $P'=\left| \psi'
\right\rangle \left\langle \psi' \right|$, where $0<\left|
\left\langle \psi|\psi^{\prime }\right\rangle \right| ^{2}<1$, we
have for this value of $\lambda$ that
$\chi_{P'}=\mathrm{Tr}(\rho_{\lambda} P')=\left| \left\langle
\psi|\psi^{\prime }\right\rangle \right| ^{2}$ and consequently
$0<\chi_{P'} (\lambda)<1$, which implies that $\chi_{P'}
(\lambda)$ is not idempotent. Thus, the assumption of
noncontextuality for unsharp measurements and outcome determinism
for sharp measurements yields a contradiction in a 2d Hilbert
space.

This no-go theorem is related to the no-go theorem of the previous
section in the same way that the no-go theorem \cite{Bell} that is
obtained from the standard Gleason's theorem \cite{Gleason} is
related to the original Kochen-Specker theorem
\cite{KochenSpecker}. The former derive a contradiction using the
full set of measurements, while the latter only make use of a
finite set.

\section{Proof of transformation contextuality in 2D}
\label{transfcontextuality}

Consider a set of six transformation procedures, denoted \textrm{T}$_{0},$%
\textrm{T}$_{\pi /3},$\textrm{T}$_{2\pi /3},$\textrm{T}$_{\pi },$\textrm{T}$%
_{4\pi /3},$\textrm{T}$_{5\pi /3},$ where the procedure \textrm{T}$_{\theta }
$ corresponds to the CP map
\begin{equation}
\mathcal{T}_{\theta }(\rho )=U_{y,\theta }\rho U_{y,\theta }^{\dag
},
\end{equation}
and where
\begin{equation}
U_{y,\theta }=\left(
\begin{array}{ll}
\cos \frac{\theta }{2} & -\sin \frac{\theta }{2} \\
\sin \frac{\theta }{2} & \cos \frac{\theta }{2}
\end{array}
\right)
\end{equation}
is the unitary operator describing a rotation by $\theta $ about
the $y$ axis in the Bloch sphere. Consider also the CP map
$\mathcal{T}$ that takes all points in the Bloch sphere and
projects them onto the $y$ axis. There are many ways of
implementing $\mathcal{T}$ as a convex sum of transformations,
specifically,
\begin{eqnarray}
\mathcal{T} &=&\frac{1}{2}\mathcal{T}_{0}+\frac{1}{2}\mathcal{T}_{\pi }
\label{K1} \\
&=&\frac{1}{2}\mathcal{T}_{\pi /3}+\frac{1}{2}\mathcal{T}_{4\pi /3}
\label{K2} \\
&=&\frac{1}{2}\mathcal{T}_{2\pi /3}+\frac{1}{2}\mathcal{T}_{5\pi /3}
\label{K3} \\
&=&\frac{1}{3}\mathcal{T}_{0}+\frac{1}{3}\mathcal{T}_{2\pi /3}+\frac{1}{3}%
\mathcal{T}_{4\pi /3}  \label{K4} \\
&=&\frac{1}{3}\mathcal{T}_{\pi /3}+\frac{1}{3}\mathcal{T}_{\pi }+\frac{1}{3}%
\mathcal{T}_{5\pi /3}  \label{K5}
\end{eqnarray}
These identities can be explained as follows. The map $\mathcal{T}$ can be
achieved by performing with probability 1/2 a rotation in the Bloch sphere
 about the $y$ axis by $
\theta $ and with probability 1/2 a rotation by $\theta +\pi$.
Taking $\theta =0,\pi /3$ and $2\pi /3$ yields Eqs.\
(\ref{K1})-(\ref{K3}). The map $\mathcal{T}$
can also be achieved by performing a rotation about $y$ by $\theta ,$ by $%
\theta +2\pi /3$ or by $\theta +4\pi /3$ with equal probabilities. Taking $%
\theta =0$ and $\pi /3$ yields Eqs.\ (\ref{K4}) and (\ref{K5}). A
rigorous proof of these statements is provided in the appendix.

By the assumption of transformation noncontextuality each of the
seven CP maps we have considered is associated with a unique
transition matrix on the space of ontic states. Suppose that we
denote the transition matrix associated with $\mathcal{T} $ by
$\Gamma,$ and the transition matrix associated with
$\mathcal{T}_{\theta }$ by $\Gamma_{\theta }.$ Because a convex
sum of transformation procedures is represented in an ontological
model by a convex sum of the associated transition matrices, Eqs.\
(\ref{K1})-(\ref{K5}) imply
\begin{eqnarray}
\Gamma &=&\frac{1}{2}\Gamma_{0}+\frac{1}{2}\Gamma_{\pi }  \label{mmm1} \\
&=&\frac{1}{2}\Gamma_{\pi /3}+\frac{1}{2}\Gamma_{4\pi /3}  \label{mmm2} \\
&=&\frac{1}{2}\Gamma_{2\pi /3}+\frac{1}{2}\Gamma_{5\pi /3}  \label{mmm3} \\
&=&\frac{1}{3}\Gamma_{0}+\frac{1}{3}\Gamma_{2\pi
/3}+\frac{1}{3}\Gamma_{4\pi /3} \label{mmm4}
\\
&=&\frac{1}{3}\Gamma_{\pi /3}+\frac{1}{3}\Gamma_{\pi
}+\frac{1}{3}\Gamma_{5\pi /3} \label{mmm5}
\end{eqnarray}

Note that $\mathcal{T}_{\theta }$ and $\mathcal{T}_{\theta +\pi }$
take any rank-1 density operator lying in the $z$-$x$ plane of the
Bloch sphere to a pair of orthogonal density operators. Since
these are distinguishable with certainty, it follows from feature
1 of ontological models (see section \ref{prepcontextuality}) that
the transition matrices $\Gamma_{\theta }$ and $\Gamma_{\theta
+\pi }$ must take any distribution $\mu_x (\lambda )$ to disjoint
distributions, that is,
\begin{equation}
\int d\lambda^{\prime}\; \Gamma_{\theta
}(\lambda,\lambda^{\prime})\;\mu_x(\lambda^{\prime})\; \int
d\lambda^{\prime}\; \Gamma_{\theta +\pi
}(\lambda,\lambda^{\prime})\;\mu_x(\lambda^{\prime})=0.
\label{orthogmaps}
\end{equation}

Now consider how our seven transition matrices affect the
distribution $ \mu _{a}(\lambda )$ associated with the density
operator $\sigma _{a}$, defined in Eq.\ (\ref{states2}) (recall
that $\sigma _{a}$ is represented on the Bloch sphere by the
vector pointing along the $z$ axis). We obtain seven distinct
distributions, which we denote $\mu _{\theta }(\lambda ) \equiv
\int d\lambda^{\prime}\; \Gamma_{\theta
}(\lambda,\lambda^{\prime})\;\mu_x(\lambda^{\prime}),$ for $\theta
=0,\pi /3,2\pi /3,\pi ,4\pi /3,5\pi /3,$ and $\mu (\lambda )
\equiv \int d\lambda^{\prime}\;
\Gamma(\lambda,\lambda^{\prime})\;\mu_x(\lambda^{\prime}).$ By
virtue of Eqs.\ (\ref{mmm1})-(\ref{mmm5}) and Eq.\
(\ref{orthogmaps}), these seven distributions satisfy Eqs.\
(\ref{c7})-(\ref{c11}) and Eqs.\ (\ref{c4})-(\ref{c6}) where
$a,A,b,B,c,C$ are associated with $\theta =0,\pi ,2\pi /3,5\pi
/3,4\pi /3,\pi /3$ respectively. But Eqs.\ (\ref{c7})-(\ref{c11})
and Eqs.\ (\ref{c4})-(\ref{c6}) cannot be satisfied
simultaneously, so we have arrived at a contradiction\footnote{It
should be noted that the above argument is equivalent to a proof
of preparation contextuality in four dimensions if one makes use
of the Jamiolkowski isomorphism between density operators in a 4d
space and CP maps in a 2d space \cite{Jamiol}.}.

In a draft of this article, the question of the existence of a
no-go theorem for transformation noncontextuality was left as an
open problem. The question was resolved by Terry Rudolph who
provided the example given above. I am grateful for his permission
to present the result here.

\section{Is the assumption of noncontextuality natural?}
\label{isNCnatural}

An important question is whether the assumption of
noncontextuality for preparations, transformations, and unsharp
measurements is \emph{as well motivated} as this same assumption
for sharp measurements, to which the notion is usually restricted.
To answer this, one must consider the motivation for the latter,
which seems to be one of ontological economy: be wary of
introducing differences in the ontological explanations of
empirical phenomena where there are no differences in the
phenomena themselves. Einstein's equivalence principle is an
example of a fruitful application of this principle. If this is
indeed the motivation, then it clearly also applies to our
generalized notions of noncontextuality. Specifically, if one
believes that equivalent statistics suggest equivalent ontological
representations for sharp measurements, why should one not believe
this for preparations, transformations, and unsharp measurements
as well? Thus, barring an alternative motivation for the
traditional notion of noncontextuality, it seems that an
ontological model that respects the statistical equivalence class
structure of preparations, transformations, and unsharp
measurements is as well (or badly) motivated as an ontological
model that respects the statistical equivalence class structure of
sharp measurements.

This of course leaves open the question of whether \emph{any}
assumption of noncontextuality is natural. The answer seems to
depend on one's interpretational bent. John Bell, for instance,
thought that contextuality was not at all surprising\footnote{Bell
states: ``The result of an observation may reasonably depend not
only on the state of the system (including hidden variables) but
also on the complete disposition of the apparatus.'' \cite{Bell}},
whereas David Mermin has characterized it as a mystery in need of
explanation\footnote{Mermin states: ``if one is attempting a
hidden variable model at all, it seems not unreasonable to expect
the model to provide the obvious explanation for this striking
insensitivity of the distribution to changes in the experimental
arrangement --- namely, that the hidden variables are
noncontextual'' \cite{Mermin}}.

In order to defend the view that measurement contextuality is
indeed mysterious within the framework of an ontological model, we
show that the reasons for thinking so are very similar to the
reasons for thinking that nonlocality is mysterious. Disregarding
classical prejudice, nonlocality is not an unreasonable
assumption. However, if the universe is fundamentally nonseparable
or is such that causal influences can propagate faster than the
speed of light, then why should it also be the case that one
cannot use these effects to achieve super-luminal signalling?
Given the presence of nonlocality at the ontological level, it
seems almost conspiratorial that one cannot make use of this
nonlocality for signalling.   Similarly, it is certainly not
unreasonable for the statistics of experimental outcomes for a
given ontic state to depend on details of the experimental
procedure. But assuming this to be the case, it is very surprising
that when one considers any valid probability distribution over
the ontic states (that is, any distribution that characterizes
what someone who knows only the preparation procedure knows about
the ontic state), the weighted average over the statistics of
outcomes does \emph{not} depend on the details of the experimental
procedure.  Again, this seems almost conspiratorial.  This analogy
suggests that removing the appearance of conspiracy from
contextuality may well be on a par with reconciling Bell's theorem
and relativity as a guide for progress in the search for a wholly
satisfactory realist interpretation of quantum theory.

It is likely that the notion of preparation noncontextuality will
also seem natural to some and unnatural to others. To shed some
light on the diversity of reactions, it is useful to distinguish
two different types of ontological model of quantum theory.
Specifically, we distinguish what we call the \emph{epistemic
view} and the \emph{ontic view} of quantum states \cite{Spekkens}.

The epistemic view of quantum states asserts that a density
operator represents nothing more than an agent's knowledge about
the ontic state of the system. Specifically, it represents the
knowledge of someone who knows only the preparation procedure. In
this view, the ontic state of a system does not fix the density
operator that is used to describe it. Distinct non-orthogonal
density operators (including the pure cases) are represented by
overlapping probability distributions within this view and are
thus consistent with a single ontic state. By contrast, the ontic
view of quantum states asserts that the density operator itself
represents an attribute of the system, and consequently that two
distinct density operators represent mutually exclusive physical
states of affairs and are therefore represented in the ontological
model by nonoverlapping (i.e. disjoint) probability distributions.

To be precise, for a set $\mathcal{S}$ of density operators
(assumed to contain some nonorthogonal elements), an ontological
model adopts an ontic view of $\mathcal{S}$ if all distinct
elements of $\mathcal{S}$ are represented by disjoint
distributions, that is,
\begin{equation}
\rho \ne \rho'  \textrm{ implies }
\mu_{\rho}(\lambda)\mu_{\rho'}(\lambda)=0 \textrm{ for all }
\rho,\rho' \in \mathcal{S},
\end{equation}
whereas an ontological model adopts an epistemic view of
$\mathcal{S}$ if only orthogonal elements of $\mathcal{S}$ are
represented by disjoint distributions
\begin{equation}
\mu_{\rho}(\lambda)\mu_{\rho'}(\lambda)=0 \textrm{ only if }
\rho\rho'=0,  \textrm{ for all } \rho,\rho' \in \mathcal{S}
\end{equation}
In other words, in an epistemic view of $\mathcal{S}$, being
orthogonal is a necessary condition for a pair of quantum states
to be represented by disjoint distributions (the argument
presented at the beginning of section \ref{prepcontextuality}
shows that orthogonality is a \emph{sufficient} condition for
disjointness, regardless of whether one adopts an ontic or an
epistemic view.)


We now show that an ontic view of the set of pure quantum states
rules out the possibility of preparation noncontextuality
\emph{trivially}.  Our purpose here is to show that an implicit
commitment to such a view can lead to the impression that the
assumption of preparation noncontextuality is unnatural.

Consider the four preparation procedures $\mathrm{P}_{a},
\mathrm{P}_{A}, \mathrm{P}_{b}$ and $\mathrm{P}_{B}$ from section
\ref{prepcontextuality}, represented in quantum theory by the
Hilbert space vectors $\psi_a,\psi_A,\psi_b$ and $\psi_B$
respectively. An ontic view of pure quantum states implies that
not only are the orthogonal states associated with disjoint
distributions,
\begin{eqnarray}
\mu_a(\lambda) \mu_A(\lambda)&=&0 \\
\mu_b(\lambda) \mu_B(\lambda)&=&0,
\end{eqnarray}
 but
also \emph{nonorthogonal} states are associated with disjoint
distributions,
\begin{eqnarray}
\mu_a(\lambda) \mu_b(\lambda)&=&0 \\
\mu_A(\lambda) \mu_b(\lambda)&=&0 \\
\mu_a(\lambda) \mu_B(\lambda)&=&0 \\
\mu_A(\lambda) \mu_B(\lambda)&=&0 \\
\end{eqnarray}
It is then clear that the preparation procedures ${\mathrm
P}_{aA}$ and ${\mathrm P}_{bB}$, obtained respectively by
implementing ${\mathrm P}_{a}$ and ${\mathrm P}_{A}$ with equal
probability, or ${\mathrm P}_{b}$ and ${\mathrm P}_{B}$ with equal
probability, are represented by distributions $\mu_{aA}$ and
$\mu_{bB}$ (defined in Eqs.~(\ref{e1}) and (\ref{e2})) that are
also disjoint,
\begin{equation}
\mu_{aA}(\lambda)\mu_{bB}(\lambda)=0.
\end{equation}
However, since these two procedures are represented by the same
density operator, namely $I/2$, they must be represented by the
same distribution in a preparation noncontextual model. Thus, an
ontic view of quantum states trivially precludes the possibility
of preparation noncontextuality.

Since our manner of speaking about pure quantum states typically
favors the ontic view of the latter, it also tends to make the
assumption of preparation noncontextuality seem implausible. The
very term ``quantum \emph{state}" already predisposes one to
thinking of the density operator as representing the physical
state of affairs rather than an agent's knowledge.  For instance,
in the context of photon polarization, the multiplicity of convex
decompositions of the completely mixed state is sometimes
summarized as follows: ``an equal mixture of states of horizontal
and vertical polarization is statistically indistinguishable from
an equal mixture of states of left and right circular
polarization''. Implicit in this sort of language is the
assumption that the four different states of polarization are
\emph{mutually exclusive} states of affairs and are therefore
ontic states. Indeed, this way of putting things compels us to
question (in vain) whether there isn't really some measurement
that \emph{could} tell these two cases apart. However, it is wrong
to take this as an argument against the ``naturalness'' of
preparation noncontextuality because this impression can be
attributed entirely to the language that is used to describe the
phenomenon.

If one is to take the epistemic view seriously, as one should in
an investigation of the possibility of an ontological model of
quantum theory, then this sort of language must be avoided, and
the assumption of preparation noncontextuality is \emph{a priori}
very plausible. Indeed, in light of the arguments that have
recently been made in favor of the epistemic view of quantum
states \cite{Ballentine,Fuchs,Emerson,Spekkens} and the fact that
one can reproduce qualitatively many quantum phenomena in
noncontextual theories
\cite{Hardydisentangling,Kirkpatrick,Spekkens,Rudolphtoy}, the
impossibility of a preparation noncontextual ontological model
appears all the more shocking to the devoted realist.

\section{The issue of outcome determinism} \label{OD}

In our proof of contextuality for unsharp measurements, we assumed
outcome determinism for sharp measurements but we assumed outcome
\emph{in}determinism for unsharp measurements. This amounts to
representing all and only those POVMs with idempotent elements by
sets of indicator functions that are idempotent. Although this
seems like a natural assumption to make, two alternative
assumptions might seem \emph{a priori }worth considering: (1) that
both sharp and unsharp measurements are outcome-deterministic, or
(2) that both are outcome-indeterministic.

We begin by considering the first alternative, that outcome
determinism also holds for \emph{unsharp }measurements. It turns
out that this is trivially inconsistent with assuming measurement
noncontextuality. Consider a measurement procedure \textrm{M
}associated with the POVM $\{I/2,I/2\}$. As argued in section
\ref{BKStheoremin2d}, the assumption of measurement
noncontextuality implies that \textrm{M} must be represented in an
ontological model by the set of indicator functions $\{1/2,1/2\}$
which are \emph{not} idempotent, and thus \textrm{M }cannot be
outcome deterministic. A recent result by Cabello \cite{Cabello}
also rules out the possibility of a hidden variable model that is
measurement noncontextual and outcome deterministic for unsharp
measurements. However, this proof is unnecessarily complex since a
consideration of the POVM $\{I/2,I/2\}$ yields the result
immediately.

The second alternative is that both sharp and unsharp measurements
are outcome-indeterministic. This is the more significant
alternative, because it constitutes the weakest assumption and
consequently the most general framework for an ontological model.
Indeed, unless the assumption of outcome determinism can itself be
justified by the assumption of noncontextuality, it is
inappropriate to call any no-go theorem that makes use of this
assumption a proof of contextuality, because in the face of a
contradiction one can always assume that the faulty assumption was
that of outcome determinism rather than that of measurement
noncontextuality. Thus, neither the proof of Bell \cite{Bell}, nor
the proof of Kochen and Specker \cite{KochenSpecker}, nor any of
the proofs of these types including those presented in section
\ref{BKStheoremin2d}, serve to rule out the possibility of
measurement noncontextuality (in the sense in which we have
defined the term). It turns out, however, that the assumption of
outcome determinism for sharp measurements can be justified by an
assumption of \emph{preparation} noncontextuality, as we shall
presently demonstrate. Given this inference, the old proofs are
vindicated insofar as they remain proofs of the impossibility of
\emph{universal} noncontextuality (noncontextuality for all
experimental procedures).

It should be noted that Toner, Bacon, and Ben-Or \cite{TonerBacon}
have considered a third alternative, namely, that outcome
determinism holds for just those POVMs with elements that are not
repeatable, that is, elements that cannot appear twice in a single
POVM, and have obtained a nontrivial no-go theorem. Bacciagaluppi
\cite{Bacciagaluppi} has considered a similar alternative and
obtained a similar result.  Although this is a much weaker
assumption than the first alternative, the resulting theorems are
still not proofs of the impossibility of universal
noncontextuality, according to our definition, since the
assumption of outcome determinism for these special POVMs has not
been justified by an assumption of universal noncontextuality. In
contrast, outcome determinism for all sharp measurements can be so
justified.  We turn now to the proof of this statement.

\subsection{Preparation noncontextuality implies outcome determinism
for sharp measurements} Consider a rank-1 PVM $\{P_{k}\}.$
Thinking of each of the elements as a rank-1 density operator,
$\rho _{k}=P_{k},$ we obtain an orthogonal set of rank-1 density
operators $\{\rho _{k}\}.$ We denote the density operators and
projectors differently because they are represented differently in
the ontological model. The set $\{\rho _{k}\}$ is represented by a
set of probability densities $\{\mu _{k}(\lambda )\},$ while the
PVM $\{P_{k}\}$ is represented by a set of indicator functions
$\{\xi _{k}(\lambda )\}.$ Since the $\rho _{k}$ are orthogonal,
the associated preparations are distinguishable with certainty,
and thus by feature 1 of ontological models we must have
\begin{equation}
\mu _{k}(\lambda )\mu _{k^{\prime }}(\lambda )=\delta _{k,k^{\prime }}.
\label{star}
\end{equation}
The support of $\mu _{k}(\lambda ),$ denoted $\Omega _{k},$ is the
region of the ontic state space assigned non-zero probability by
$\mu
_{k}(\lambda ),$%
\begin{equation}
\label{starstar}
\Omega _{k}=\{\lambda |\mu _{k}(\lambda )>0\}.
\end{equation}
Eq.\ (\ref{star}) then implies that
\begin{equation}
\Omega _{k}\cap \Omega _{k^{\prime }}=\emptyset \text{ if }k\ne k^{\prime }.
\end{equation}
\strut

\strut Now, by virtue of the fact that
\begin{equation}
Tr(\rho _{k}P_{k^{\prime }})=\delta _{k,k^{\prime }},
\end{equation}
we infer that
\begin{equation}
\int \xi _{k}(\lambda )\mu _{k^{\prime }}(\lambda )=\delta _{k,k^{\prime }}.
\end{equation}
But, given Eq.\ (\ref{starstar}), this implies that
\begin{equation}
\xi _{k}(\lambda )=
\begin{array}{l}
1\text{ for }\lambda \in \Omega _{k} \\
0\text{ for }\lambda \in \cup _{k^{\prime }\ne k}\Omega _{k^{\prime }}
\end{array}
,
\end{equation}
or, equivalently,
\begin{equation}
\xi _{k}(\lambda )\xi _{k^{\prime }}(\lambda )=\delta _{k,k^{\prime }}\text{
for }\lambda \in \cup _{j}\Omega _{j}.  \label{star2}
\end{equation}
So, if one can show that the union of the supports of the $\mu
_{k}(\lambda ) $ is the entire ontic state space, i.e.,
\begin{equation}
\cup _{j}\Omega _{j}=\Omega ,  \label{star3}
\end{equation}
then Eq.\ (\ref{star2}) would imply that $\{\xi _{k}(\lambda )\}$
is a set of idempotent indicator functions, and consequently would
establish that our rank-1 PVM must be outcome-deterministic in the
ontological model.

It turns out that Eq.\ (\ref{star3}) follows from the assumption
of preparation noncontextuality. First note that the ontic state
space $\Omega $ can be defined as the set of $\lambda $ that are
assigned non-zero probability by \emph{some} density operator
\begin{equation}
\Omega =\left\{ \lambda |\mu _{\rho }(\lambda )>0\text{ for some }\rho
\right\} .
\end{equation}
However, since every density operator $\rho $ appears in some
convex decomposition of the completely mixed state $I/d$ (where
$d$ is the dimensionality of the Hilbert space), and since
preparation noncontextuality implies that there is a unique
distribution $\mu_{I/d}(\lambda)$ associated with this state, it
follows that $\Omega $ is simply the set of $\lambda $ assigned
non-zero probability by the latter, i.e.,
\begin{equation}
\Omega =\left\{ \lambda |\mu _{I/d}(\lambda )>0\right\} .
\end{equation}
But given that the $\rho _{k}$ form a convex decomposition of $I/d,$
\begin{equation}
\sum_{k}\frac{1}{d}\rho _{k}=\frac{I}{d},
\end{equation}
it follows from preparation noncontextuality that
\begin{equation}
\sum_{k}\frac{1}{d}\mu _{k}(\lambda )=\mu _{I/d}(\lambda ),
\end{equation}
which implies Eq.\ (\ref{star3}).

This establishes outcome determinism for PVMs all of whose
elements are rank 1. Since an arbitrary PVM can always be obtained
by coarse-graining of a rank-1 PVM, and since coarse-graining
takes idempotent functions to idempotent functions, \emph{any} PVM
is represented by a set of idempotent indicator functions. This
establishes that the assumption of outcome determinism for sharp
measurements follows from an assumption of preparation
noncontextuality.

It is natural to wonder whether outcome determinism for sharp
measurements might be justified by an assumption of measurement
noncontextuality (rather than an assumption of preparation
noncontextuality). If this were possible, then the proofs in
section \ref{BKStheoremin2d} would derive contradictions from
measurement noncontextuality alone. It turns out that this is not
possible, because measurement noncontextuality on its own is
consistent with quantum theory, as we now show.

\subsection{Achieving measurement noncontextuality by giving up
outcome determinism} \label{achievingmeasNC}

Consider the following ontological model of quantum theory, which
is objectively indeterministic and adopts an ontic view of quantum
states. The ontic state space $\Omega $ is simply taken to be the
projective Hilbert space, that is, the set of rays of Hilbert
space. Thus, for every rank-1 projector $\left| \psi \right\rangle
\left\langle \psi \right| ,$ we associate a single ontic state,
which we denote by $\psi $. Consequently, there are no hidden
variables in this ontological model. A preparation procedure
associated with the rank-1 density operator $\left| \psi ^{\prime
}\right\rangle \left\langle \psi ^{\prime }\right| $ is
represented by a Dirac-delta distribution
\begin{equation}
\mu _{\psi ^{\prime }}(\psi )=\delta (\psi -\psi ^{\prime }).
\end{equation}
A preparation procedure involving a convex combination of rank-1
density operators $ \{p(\psi'),\left| \psi'\right\rangle
\left\langle \psi'\right| \}$ is represented by the distribution
\begin{equation}
\mu (\psi )=\int d\psi' p(\psi') \delta (\psi -\psi'),
\label{mixture2}
\end{equation}
where $d\psi$ is the unitarily-invariant measure on the projective
Hilbert space. A measurement of the POVM $\{Q_{k}\}$ is associated
with a set of indicator functions $\{\xi _{Q_{k}}(\lambda )\}$
defined by
\begin{equation}
\xi _{Q_{k}}(\psi )=\mathrm{Tr}(Q_{k}\left| \psi \right\rangle
\left\langle \psi \right| ).
\end{equation}
These functions are clearly positive by virtue of the positivity
of the $Q_{k},$ and sum to unity by virtue of the fact that
$\sum_{k}Q_{k}=I.$ Note also that they depend only on the POVM
that is associated with the measurement and not on how it was
implemented. One can see that this model reproduces quantum theory
by noting that
\begin{equation}
\int \mu _{\psi ^{\prime }}(\psi )\xi _{Q_{k}}(\psi )d\psi
=\mathrm{Tr}(Q_{k}\left| \psi ^{\prime }\right\rangle \left\langle
\psi ^{\prime }\right| ).
\end{equation}
The predictions for mixed preparations are also reproduced.

This model has been discussed at length by Beltrametti and
Bugajski \cite {BeltramettiBugajski}, and captures to some extent
the ontological model that many physicists implicitly adhere to.
Note that the model is obviously preparation \emph{contextual}
since the distribution that represents a convex combination of
preparation procedures, described in Eq.\ (\ref{mixture2}),
depends on the particular ensemble of pure states, and not just on
the density operator associated with the mixture. This fact comes
as no surprise since the results of section
\ref{prepcontextuality} show that \emph{any} ontological model,
deterministic or not, must be preparation contextual. More
importantly for the purposes of this section, the set of indicator
functions associated with any PVM $\{P_k\}$ are not idempotent.
This is clear since $\mathrm{Tr}(P_{k}\left| \psi\right\rangle
\left\langle \psi \right| )$ is only 0 or 1 if $\left|
\psi\right\rangle$ lies in an eigenspace of $P_k$. It follows that
the assumption of outcome determinism for sharp measurements is
explicitly violated. However, because the set of indicator
functions depends only on the POVM, and not on its context, the
assumption of measurement noncontextuality is upheld.

\section{Conclusions}
\label{conclusions}

Because the traditional notion of noncontextuality only allowed
for a no-go theorem in Hilbert spaces of dimensionality greater
than two, there have been many proposed hidden variable models for
2d Hilbert spaces that are purported to be noncontextual
\cite{KochenSpecker,Bell}. These have been presented primarily as
pedagogical examples of what sort of model is excluded for
larger-dimensional Hilbert spaces. However, by our generalized
definition of noncontextuality, all of these models are deemed
\emph{contextual} by virtue of being contextual for preparations,
transformations, and unsharp measurements. This overturns the
notion, suggested by the restriction of old Bell-Kochen-Specker
theorems to Hilbert spaces of dimensionality greater than two,
that there is nothing inherently nonclassical about a 2d Hilbert
space \cite{vanEnk}.

In the face of this claim, a sceptic might argue that the proofs
presented here have made use of mixed preparations, unsharp
measurements, and irreversible transformations (associated
respectively with non-rank-1 density operators, non-projective
POVMs, and non-unitary CP maps), and that these are necessarily
implemented in practice through pure preparations, sharp
measurements, and reversible transformations (associated
respectively with rank-1 density operators, PVMs, and unitary
maps) on a larger system and therefore implicitly make use of a
Hilbert space of dimension greater than two. However, this is
incorrect. If one examines carefully the proofs presented in this
article, one finds that wherever non-rank-1 density operators,
non-projective POVMs, or non-unitary CP maps arise, they are due
to ignorance of which of several rank-1 density operators, PVMs,
or unitary maps in the 2d Hilbert space is appropriate, rather
than being due to the neglect of a subspace or subsystem of a
larger dimensional Hilbert space. In other words, any
``ancillary'' systems used to implement such procedures can be
treated classically, and thus do not require one to posit a larger
Hilbert space.

Our operational definition of noncontextuality has allowed us to
distinguish the notions of preparation, transformation, and
measurement noncontextuality. Our proof of preparation
contextuality is particularly novel as a no-go theorem insofar as
it focusses on the impossibility of reproducing, within a
particular kind of ontological model, the \emph{convex structure
of the set of quantum states} rather than the algebraic structure
of the set of quantum measurements.  It is interesting to note
that wherever one finds a freedom of decomposition in the
formalism of operational quantum theory, such as the multiplicity
of convex decompositions of a mixed quantum state or of a POVM
element, the multiplicity of fine-grainings of a non-rank-1 POVM,
or the unitary freedom in the operator-sum representation of a
non-unitary CP map, one can develop a proof of contextuality that
is based on this freedom.

We have shown that one can confine all the contextuality into the
preparations and transformations if one likes, because there exist
outcome-indeterministic ontological models of quantum theory, such
as the Beltrametti-Bugajski model, that are measurement
noncontextual. On the other hand, one cannot confine all the
contextuality into the measurements, because the assumption of
preparation noncontextuality yields a contradiction on its own. In
this sense, preparation contextuality is more fundamental to
quantum theory than measurement contextuality.

The issue of noncontextuality is closely linked with the issue of
locality. Indeed, it is sometimes claimed that nonlocality is an
instance of measurement contextuality.  If this were the case,
then proofs of nonlocality would also constitute proofs of
measurement contextuality, and since there exist proofs of
nonlocality that do not assume outcome determinism for sharp
measurements, it would appear that there should exist proofs of
measurement contextuality that do not make this assumption either.
But this would be in contradiction with the claims of the previous
section.

The resolution of this puzzle is that one can distinguish two
sorts of locality \cite{DonHoward}, and it is only the failure of
one of these that implies measurement contextuality. The first
notion of locality, which we call \emph{separability}, is the
assumption that the ontic state of the universe is defined in
terms of the ontic states at each point of space-time.  The other
sort of locality assumption, which presumes separability, we call
\emph{local causality}.  It is the assumption that the probability
distribution over values for a variable in a space-time region are
determined by the values of all the variables in the backward
light-cone of this region (see footnote in section
\ref{NCforquantumtheory}). A failure of local causality within the
framework of a separable model \emph{does} indeed imply
measurement contextuality. However, a model can be nonlocal by
virtue of failing to be separable, and in this case it does not
follow that the model is measurement contextual. This is precisely
what occurs in the Beltrametti-Bugajski model. The variables for a
composite system are not simply the Cartesian product of the
variables of the components, since the Cartesian product of two
projective Hilbert spaces is not the projective Hilbert space of
the tensor product (it fails to include the entangled states). In
particular, spatially separated systems are not associated with
distinct variables. Thus, the Beltrametti-Bugajski model is not
separable. It is only within the context of a separable theory
that Bell's theorem implies measurement contextuality.

The opposite inference has also received a great deal of
attention: whether a proof of measurement contextuality can be
turned into a proof of nonlocality (note that the question is only
interesting if one presumes separability since otherwise one is
already acknowledging a failure of some sort of locality). The
motivation for this investigation is clear since, as Bell famously
emphasized, an assumption of measurement noncontextuality is most
compelling if it can be justified by an assumption of locality
\cite{Bell2}. Many authors have shown how certain no-go theorems
for measurement noncontextuality can be turned into no-go theorems
for locality by virtue of the fact that sometimes every assumption
of measurement noncontextuality in a Bell-Kochen-Specker theorem
can be justified by an assumption of locality
\cite{Mermin,Penrose}. It turns out that the same trick can be
achieved in no-go theorems for preparation noncontextuality.
Although the particular proof of the no-go theorem presented in
section \ref{prepcontextuality} does not admit such a
justification, a proof can be found which does. This will be
presented in a separate article \cite{RudolphSpekkens}. The
version of Bell's theorem that results is particularly
enlightening, as it constitutes a more direct response to the EPR
argument \cite{EPR} compared to standard versions of the theorem.

It should be noted that there are contexts that do not have any
representation in the formalism of operational quantum theory.
Whether one uses a piece of polaroid or a birefringent crystal in
a measurement of photon polarization is an example of such a
context. No dependence on \emph{this sort of context} is implied
by any of the no-go theorems we have presented\footnote{If,
however, one decides to treat part of the experimental apparatus
as a quantum system, then this sort of distinction \emph{could} be
represented within the quantum formalism and it is then
conceivable that one could prove that context-dependence of this
type is sometimes required.}. Nonetheless, some hidden variable
theories still exhibit such dependence. For instance, it has been
shown that the deBroglie-Bohm interpretation has this sort of
context-dependence for certain position measurements
\cite{Valentini,Hardy95} and for certain spin measurements
\cite{Albert}. Thus, the deBroglie-Bohm interpretation involves
more contextuality than has been shown to be required of an
ontological model. Note that Refs.\ \cite{Valentini, Hardy95,
Albert} explicitly identify this feature of the deBroglie-Bohm
interpretation as a kind of contextuality despite the fact that it
does not fit into the standard definition of contextuality
presented in the introduction.  The possibility of this type of
phenomenon was in fact considered in the framework of a general
hidden variable theory much earlier by Shimony \cite{Shimony}, who
also described it as a kind of contextuality. This highlights
another virtue of our generalized definition of contextuality: it
accords with the intuition that a measurement context is
\emph{any} feature of the measurement that is not specified by
specifying its equivalence class.

An operational definition of noncontextuality is also likely to be
useful because it allows one to investigate the possibility of
finding ways of experimentally differentiating the set of
noncontextual theories from the set of contextual theories, much
as the Bell inequalities differentiate all local realistic
theories from their alternatives. If these investigations are
successful, they could shed light on the question of how to
perform experimental tests of contextuality, a subject of much
recent interest \cite{Zeilinger,Larsson,CabelloGarcia}. The
question of whether an experimental test of contextuality is even
\emph{possible} has been the subject of some controversy, due to
the finite precision of real experimental procedures \cite
{Meyer,CliftonKent,Appleby,KentBarrett}.  The problem, from the
perspective of this article, is that finite precision might imply
that in practice no two experimental procedures are found to be
operationally equivalent, in which case the assumption of
noncontextuality is never applicable.  A possible resolution of
this finite precision loophole is to further generalize the
definition of noncontextuality proposed in the introduction as
follows:
\begin{quote}
A noncontextual ontological model of an operational theory is one
wherein if two experimental procedures are operationally similar,
then they have similar representations in the ontological model.
\end{quote}
To be substantive, this proposal must be supplemented by a
quantitative measure of similarity in the space of operational
procedures, and a corresponding measure in the space of
ontological representations of these procedures. Whether this
strategy can lead to an experimentally robust notion of
contextuality is a subject for future research.

Finally, given the fact that some quantum information processing
protocols, namely, protocols for communication complexity problems
\cite{Zukowski}, have been proven to require violations of Bell
inequalities in order to outperform their classical counterparts,
it is interesting to investigate whether the power of any quantum
information processing protocols might be attributed to the
contextuality of quantum theory. There is already some evidence to
this effect in the case of random access codes \cite{Galvao}. We
speculate that this might also be the case for the exponential
speed-up of a quantum computer relative to a classical computer,
if such a speed-up exists.
\medskip

\begin{acknowledgements}

I would like to thank John Sipe, who was instrumental in
developing the operational definition of noncontextuality, Terry
Rudolph for his many improvements to the contents of this paper,
Chris Fuchs, Lucien Hardy and Ben Toner for helpful discussions,
and Guido Bacciagaluppi, Joseph Emerson, Jerry Finkelstein and
Ernesto Galv\~{a}o for insightful comments on previous drafts of
the article.

\end{acknowledgements}

\appendix

\section{Proof of Eqs.~(\ref{K1})-(\ref{K5})}

To demonstrate Eqs.~(\ref{K1})-(\ref{K5}), we make use of the fact
that there is unitary freedom in the operator-sum representation
of a CP map \cite{NielsenChuang}. Suppose that $\{W_{\mu }\}$ are
a set of operators (called \emph{Kraus operators}) appearing in an
operator-sum representation of $ \mathcal{T},$ that is,
\begin{equation}
\mathcal{T}(\rho )=\sum_{\mu }W_{\mu }\rho W_{\mu }^{\dag }.
\end{equation}
Then, for any unitary matrix $u_{\nu \mu },$ the set of operators $\{X_{\nu
}\}$ defined by
\begin{equation}
X_{\nu }=\sum_{\mu }u_{\nu \mu }W_{\mu }  \label{unitaryfreedom}
\end{equation}
also forms an operator-sum representation of $\mathcal{T}.$ Note
that we allow Kraus operators to be zero, so that different
operator-sum representations may have different cardinality.

Eq. (\ref{K1}) implies that $\mathcal{T}$ has an operator-sum representation
in terms of the set of Kraus operators $\{W_{1},W_{2}\}=\{\frac{1}{\sqrt{2}}%
U_{0},\frac{1}{\sqrt{2}}U_{\pi }\},$ since
\begin{equation}
\mathcal{T}(\rho )=\frac{1}{2}U_{0}\rho U_{0}^{\dag }+\frac{1}{2}U_{\pi
}\rho U_{\pi }^{\dag }.
\end{equation}
The set of operators $\{X_{1},X_{2}\}=\{\frac{1}{\sqrt{2}}U_{\theta },\frac{1%
}{\sqrt{2}}U_{\theta +\pi }\}$ also yield an operator-sum representation of $%
\mathcal{T}$ since they can be obtained by a unitary remixing of
$\{W_{1},W_{2}\}$, via Eq.~(\ref{unitaryfreedom}), using the $2
\times 2$ unitary matrix
\begin{equation}
u=\left(
\begin{array}{ll}
\cos \frac{\theta }{2} & \sin \frac{\theta }{2} \\
-\sin \frac{\theta }{2} & \cos \frac{\theta }{2}
\end{array}
\right)
\end{equation}
It follows, in particular, that the sets of Kraus operators $\{\frac{1}{%
\sqrt{2}}U_{\pi /3},\frac{1}{\sqrt{2}}U_{4\pi /3}\}$ and $\{\frac{1}{\sqrt{2}%
}U_{2\pi /3},\frac{1}{\sqrt{2}}U_{5\pi /3}\}$ form operator-sum
representations of $\mathcal{T},$ and consequently that Eqs. (\ref{K2}) and (%
\ref{K3}) hold.

Next, we show that the set of operators $\{X_{1},X_{2},X_{3}\}=\{\frac{1}{%
\sqrt{3}}U_{\theta },\frac{1}{\sqrt{3}}U_{\theta +2\pi /3},\frac{1}{\sqrt{3}}%
U_{\theta +4\pi /3}\}$ also yield an operator-sum representation of $%
\mathcal{T}.$ First note that the set $\{W_{1},W_{2},W_{3}\}=\{\frac{1}{%
\sqrt{2}}U_{0},\frac{1}{\sqrt{2}}U_{\pi },0\}$ yields the operator-sum
representation of $\mathcal{T}$ associated with Eq. (\ref{K1}). The
operators $\{X_{1},X_{2},X_{3}\}$ can be obtained by a unitary remixing of $%
\{W_{1},W_{2},W_{3}\}$ using the 3$\times 3$ unitary matrix
\begin{equation}
u=\left(
\begin{array}{lll}
\sqrt{\frac{2}{3}}\cos \frac{\theta }{2} & \sqrt{\frac{2}{3}}\sin \frac{%
\theta }{2} & \sqrt{\frac{1}{3}} \\
\sqrt{\frac{2}{3}}\cos (\frac{\theta }{2}+\frac{2\pi }{3}) & \sqrt{\frac{2}{3%
}}\sin (\frac{\theta }{2}+\frac{2\pi }{3}) & \sqrt{\frac{1}{3}} \\
\sqrt{\frac{2}{3}}\cos (\frac{\theta }{2}+\frac{4\pi }{3}) & \sqrt{\frac{2}{3%
}}\sin (\frac{\theta }{2}+\frac{4\pi }{3}) & \sqrt{\frac{1}{3}}
\end{array}
\right) .
\end{equation}
It follows, in particular, that $\{\frac{1}{\sqrt{3}}U_{0},\frac{1}{\sqrt{3}}%
U_{2\pi /3},\frac{1}{\sqrt{3}}U_{4\pi /3}\}$ and $\{\frac{1}{\sqrt{3}}U_{\pi
/3},\frac{1}{\sqrt{3}}U_{\pi },\frac{1}{\sqrt{3}}U_{5\pi /3}\}$ form
operator-sum representations of $\mathcal{T}$ and consequently that Eqs. (\ref{K4}) and (\ref{K5}) hold.

\end{document}